\documentclass[twocolumn,11pt]{revtex4-1} 
\usepackage{amsmath,amssymb,graphicx,verbatim,setspace}
\begin{document}

\title{Epoxy-based broadband anti-reflection coating for millimeter-wave optics}

\author{Darin Rosen,$^{1,*}$ Aritoki Suzuki,$^{1,*}$ Brian Keating,$^2$ William Krantz,$^1$ Adrian T. Lee,$^1$ Erin Quealy,$^1$ Paul L. Richards,$^1$ Praween Siritanasak,$^2$ and William Walker$^1$}
\address{$^1$Department of Physics, University of California, Berkeley, \\ 278 LeConte Hall Berkeley, CA, 94720, USA}
\address{$^2$Department of Physics, University of California, San Diego, \\ 9500 Gilman Drive La Jolla, CA, 92092, USA}
\address{$^*$Corresponding authors: darosen@berkeley.edu, asuzuki@berkeley.edu}

\begin{spacing}{1.0}
\begin{abstract}
We have developed epoxy-based, broadband anti-reflection coatings for millimeter-wave astrophysics experiments with cryogenic optics. By using multiple-layer coatings where each layer steps in dielectric constant, we achieved low reflection over a wide bandwidth. We suppressed the reflection from an alumina disk to 10\% over fractional bandwidths of 92\% and 104\% using two-layer and three-layer coatings, respectively. The dielectric constants of epoxies were tuned between 2.06 and 7.44 by mixing three types of epoxy and doping with strontium titanate powder required for the high dielectric mixtures. At 140~Kelvin, the band-integrated absorption loss in the coatings was suppressed to less than 1\% for the two-layer coating, and below 10\% for the three-layer coating. 
\end{abstract}


\maketitle 

\section{Introduction}\label{sec:intro}
\vspace{-15pt}
Alumina and silicon are ideal lens materials for cryogenic millimeter-wave optics due to their high thermal conductivities and high dielectric constants. A high thermal conductivity facilitates cooling of a lens to reduce thermal emission and absorption loss while a high dielectric constant ($\epsilon_r \approx10-12$) allows for the fabrication of thin lenses with small radii of curvature. However, one downside of the high dielectric constant is reflection at the vacuum-dielectric interface, which can be as high as 30\%. There are many effective anti-reflection (AR) coatings using a thin dielectric coating, metal-mesh layers or sub-wavelength structures \cite{Lau,Quealy,Zhang,McMahon}. However, these coatings will not suffice for some of the next generation Cosmic Microwave Background (CMB) polarimetry experiments, such as POLARBEAR-2 \cite{Tomaru}, a broadband experiment that utilizes 50 cm diameter alumina lenses and 5.345 mm diameter hemispherical silicon lenses to focus light on the detectors. It is challenging to machine sub-wavelength structures in alumina due to its hardness. Metal-mesh coatings are not available in a 50 cm diameter size. Most reported millimeter-wave, dielectric-based AR coatings are a single layer and thus limited to narrow bandwidths. A single-layer coating for our application would have a 41\% fractional bandwidth with under 10\% reflection. We have developed multilayer epoxy-based dielectric AR coatings with more than 90\% fractional bandwidth. While multilayer dielectric coatings have been developed \cite{Quealy,EBEX1,EBEX2}, our moldable adhesive coatings have the advantage of being applicable for high dielectric constant curved lenses. Additionally, we can tune the dielectric constant of our layers, which allows for broad application of our coatings.\\
\indent An analytical solution for wave propagation in multiple layers of dielectrics can be obtained with the characteristic matrix method \cite{Heavens}. In general, the bandwidth of coatings increases with the number of correctly tuned layers, but the absorptive loss also increases due to increased thickness. Therefore, we optimized the dielectric constant and thickness to keep the number of layers as small as possible and still provide large bandwidth. For millimeter-wave application, the thickness of the AR coating will be \textit{O}(100~$\mu\mathrm{m}$). Thus the thickness must be controlled to \textit{O}(10~$\mu\mathrm{m}$). We would also like to be able to tune the dielectric constant of the coating material to obtain optimal performance.
For a ground-based CMB experiment, we require approximately 30\% bandwidth around 95~GHz, 150~GHz and 220~GHz corresponding to atmospheric windows. We calculate that a two-layer coating with dielectric constants of $\epsilon=$ 2 and 5 with thicknesses of one-quarter wavelength at 120~GHz will give sufficient bandwidth to cover both 95~GHz and 150~GHz bands. To cover 95, 150 and 220~GHz bands simultaneously, we propose a three-layer coating with dielectric constants of $\epsilon=$ 2, 4, and 7 with thickness of a quarter of wavelength at 150~GHz.
\vspace{-15pt}
\section{Coating Material}\label{sec:material}
\begin{figure}[htbp]
\centering
\includegraphics[width=8.0cm]{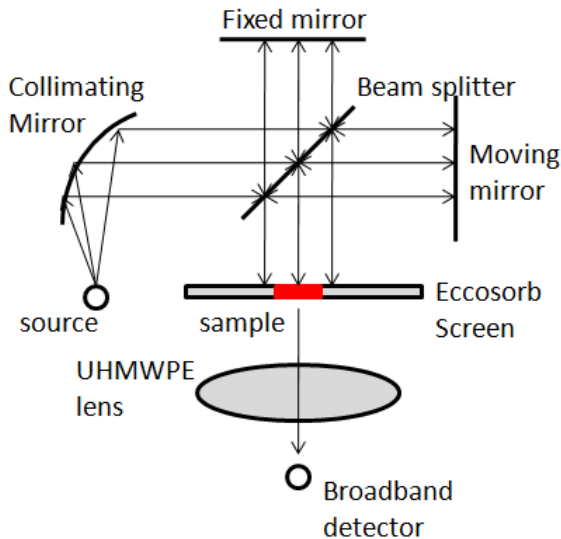}
\vspace{-7pt}
\caption{A schematic of the Michelson FTS measurement. We placed the sample at the collimated output of the FTS. An absorber (eccosorb AN-72) was placed around the aperture. The signal was collimated by an UHMWPE lens to a broadband (70-250 GHz) bolometric detector.}
\label{fig:FTS}
\end{figure}

We chose epoxy as the base material for its adhesion properties and its ability to be molded  into any shape. We referred to Lamb for the approximate dielectric constants of epoxies \cite{Lamb}. To measure dielectric constants and absorption losses, we used the Michelson Fourier transform spectrometer (FTS) measurement shown in Figure~\ref{fig:FTS}. 

The FTS uses the contrast between an 800~Kelvin ceramic heater and a 300~Kelvin Eccosorb AN-72 absorber. Mirrors are 152 x 152 mm in cross-section. The beam splitter is made of 0.25 mm thick Mylar which has a peak efficiency at 180~GHz. The sample holder, which is placed at the output of the FTS, has a 51 mm diameter aperture. An absorbing screen terminates rays that do not go through the aperture. The rays that go through the aperture are focused onto a broadband detector using an ultra high molecular weight polyethylene (UHMWPE) lens. For the detector, we used a broadband antenna-coupled transition-edge sensor bolometer with superconducting quantum interference device (SQUID) readout\cite{o'brient:063506,Suzuki} and observed frequencies between 70~GHz to 250~GHz with a resolution of 1.6~GHz. 

\begin{figure}
\centering
\includegraphics[width=8.4cm]{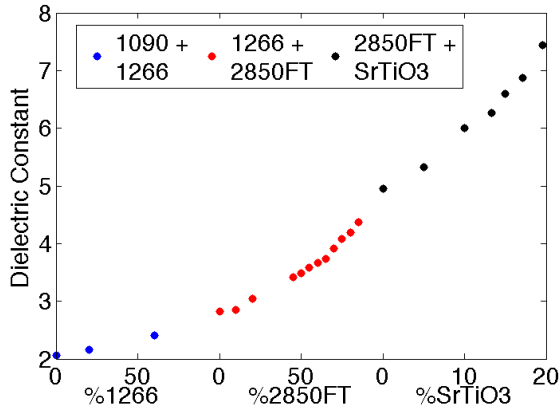}
\vspace{-7pt}
\caption{Dielectric constants of various epoxy and $\mathrm{SrTiO_3}$ mixtures at room temperature as a function of the percent by weight of the total mixture.}
\label{fig:epoxy}
\end{figure}

We mixed Emerson and Cuming's Stycast 1090, Stycast 1266A and Stycast 2850FT with their corresponding catalysts, Catalyst 9, Stycast 1266B and Catalyst 23LV respectively. For the mixing ratio we followed each product's data sheets and avoided mixing more than 100~mL of sample at a time as heat from the exothermic reaction hardens the mixture too quickly for our application.  Cylindrical aluminum molds 25 mm deep and 51 mm in diameter were coated with Mann Ease Release 200 mold release. We poured the mixture into the mold, then placed the mold in a $90^\circ\mathrm{C}$ oven for a few hours. 
We cut the cured samples to approximately 6~mm thick and machined both sides to be parallel within 0.1~mm which is approximately $\lambda/20$ depending on the exact frequency and refractive index of the sample. We finished the surface of the sample with 400 grit sand paper.
To obtain the transmittance properties of a given sample, we divided a spectrum with the sample in the beam by a spectrum without the sample. We averaged three pairs of such data. Fabry-Perot fringes in transmittance data occur with frequency spacing $\Delta f = c/(2dn)$ where $d$ is the thickness of a sample. We found that Stycast 1090, Stycast 1266A and Stycast 2850FT have dielectric constants of 2.06, 2.60 and 4.95 respectively and successfully obtained mixtures with intermediate dielectric constants by mixing two types of epoxy. To obtain a dielectric constant higher than 4.95, we mixed Stycast 2850FT with $\mathrm{SrTiO_3}$ powder from Fisher Scientific which has been shown to have a high dielectric constant at lower frequency\cite{Lee}. For these high dielectric mixtures, we vacuum-pumped the mixture for 5 minutes to remove air bubbles.  With a Stycast 2850FT and $\mathrm{SrTiO_3}$ mixture, we obtained dielectric constants as high as 7.44. We summarize the results in Figure~\ref{fig:epoxy}. 

\section{Anti-Reflection Coating}\label{sec:arcoating}
\begin{figure}[htbp]
\centering
\includegraphics[width=8.4cm]{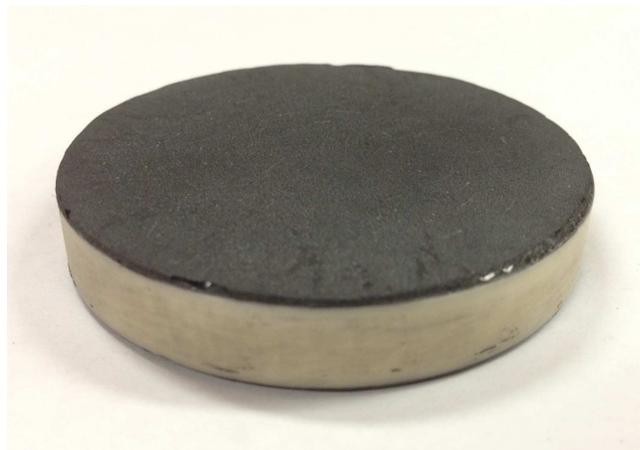}
\vspace{-7pt}
\caption{Photograph of three-layer AR coated alumina. The alumina sample is 51 mm in diameter and 6.35 mm thick.}
\label{fig:sample}
\end{figure}

We AR-coated cylindrical 99.5\% purity alumina samples from Coorstek which have a dielectric constant of 9.6. For better adhesion, we lightly sanded the surface of the alumina sample with 400 grit sand paper prior to applying the coatings. We prepared epoxy mixtures as described in Section~\ref{sec:material} and then applied a thin layer of the mixture on the alumina sample. After the mixture cured, we sanded down each layer to 25~$\mu\mathrm{m}$ thickness accuracy before applying next layer. For an expedited curing process, we placed samples in a $90^\circ \mathrm{C}$ oven for a few hours. However, the highest dielectric layer with Stycast 2850FT and $\mathrm{SrTiO_3}$ was particularly difficult to sand when fully cured. For a coating with this mixture, we removed the sample from the oven after 45 minutes to sand the surface before it had completely cured. A finished sample is shown in Figure~\ref{fig:sample}.

\begin{figure}[htbp]
\centering
\includegraphics[width=8.4cm]{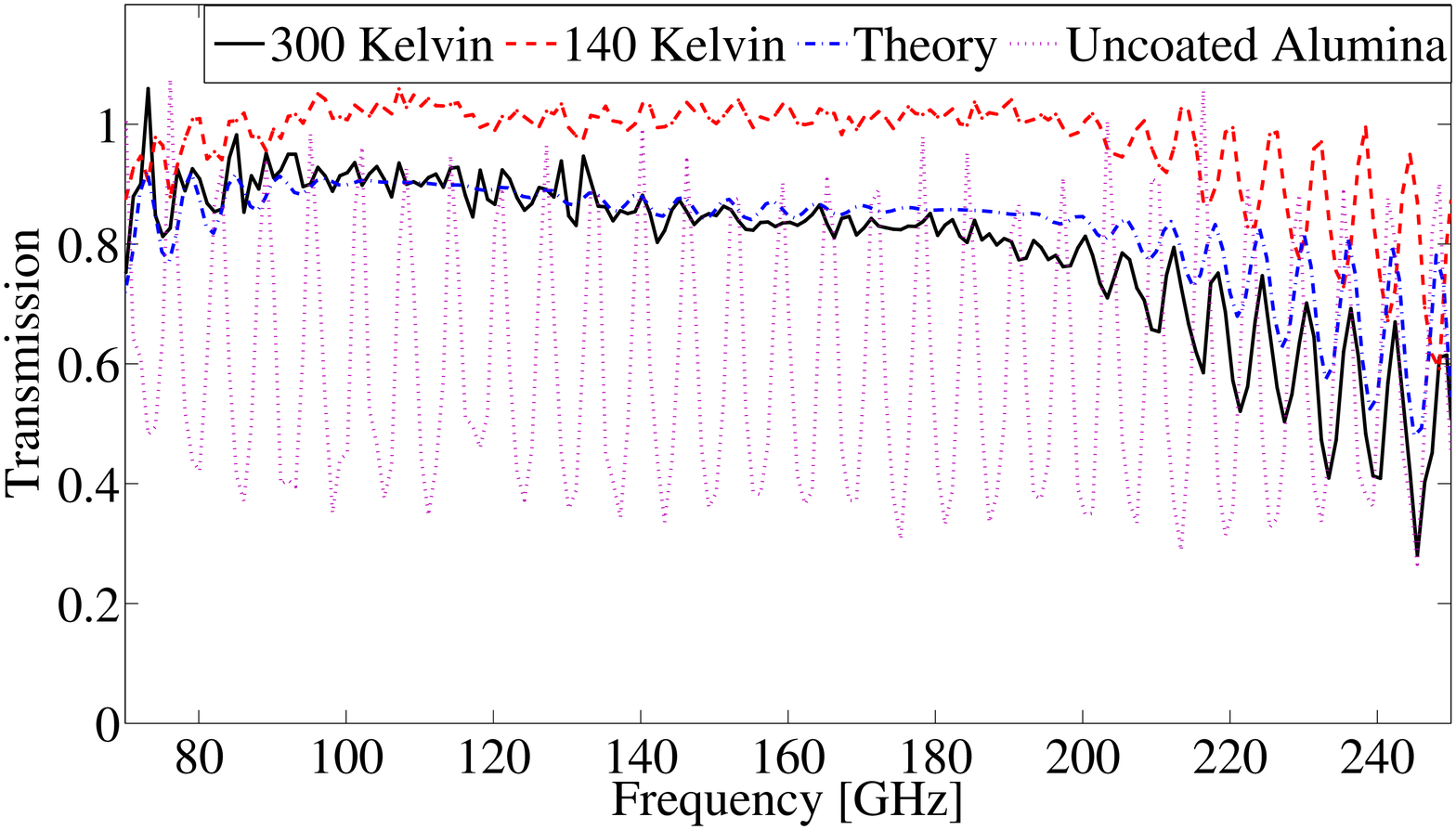}
\vspace{-10pt}

\label{fig:2ar}

\centering
\includegraphics[width=8.4cm]{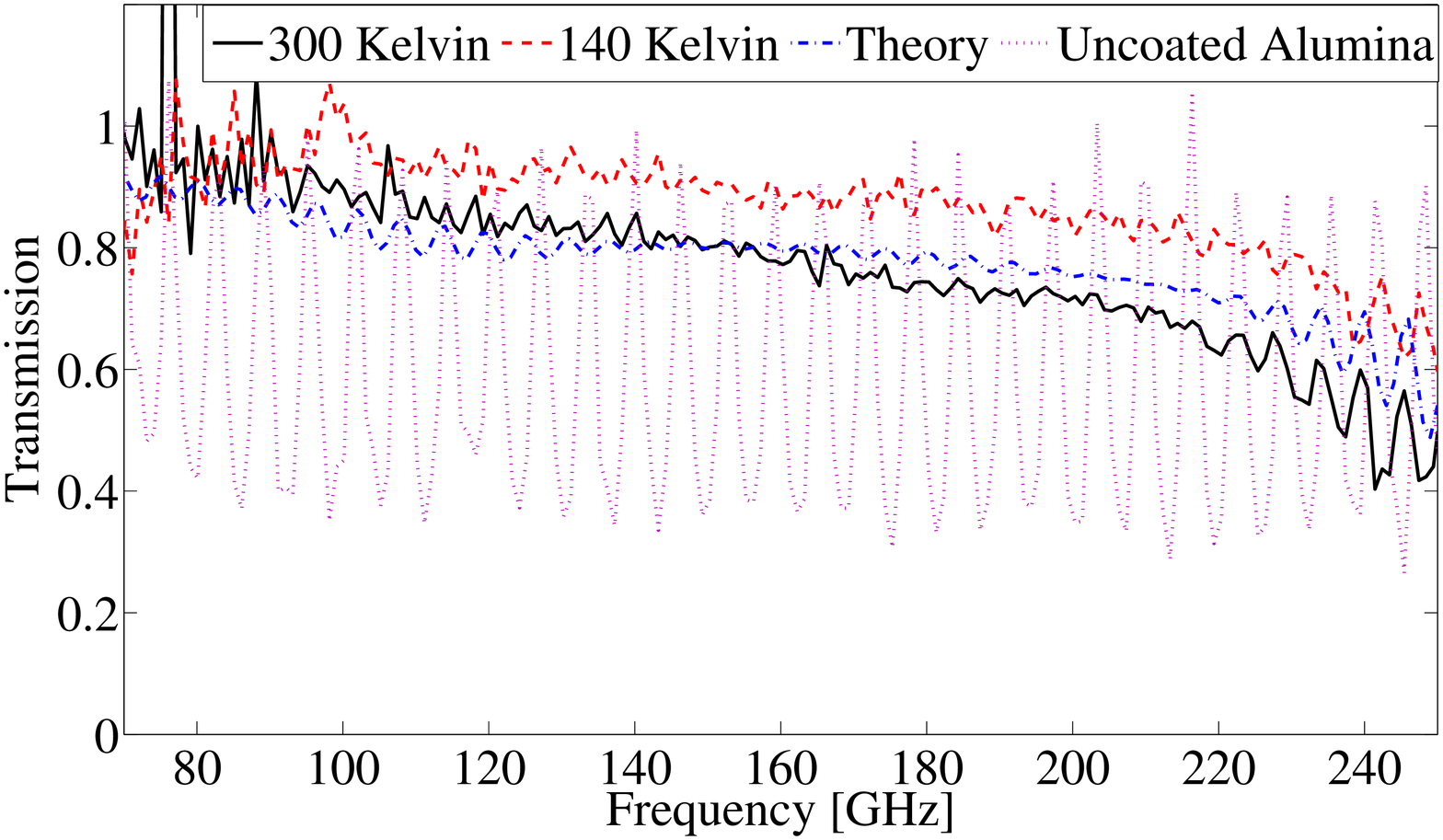}
\vspace{-20pt}
\caption{Transmittance spectra of two-layer (top) and three-layer (bottom) AR coated alumina at 300 Kelvin (solid black) and 140 Kelvin (dashed red), the modeled curve at 300 Kelvin (dash-dotted blue), and uncoated alumina (dotted magenta). A widened transmittance band can be inferred from the lack of Fabry-Perot fringes.}
\label{fig:3ar}
\end{figure}

We measured room temperature performance using the FTS described in Section~\ref{sec:material}. Transmittance plots for two-layer and three-layer coatings are shown in Figures~\ref{fig:2ar} and~\ref{fig:3ar} respectively. The measurement shows uncoated alumina which has high Fabry-Perot fringes due to high reflection, whereas the coated sample has high transmittance over a wide band. The measured and modeled curves agree well, however with the modeled curve, we assumed a constant loss-tangent at 150~GHz. The slight mismatch between the data and theory comes from an increase in loss-tangent as function of frequency, which is a typical loss trend for epoxy in the millimeter range \cite{Lamb}.

We cooled the sample to 140~Kelvin by conduction with a copper sample holder partially immersed in liquid nitrogen. This setup allows the sample to stay above the liquid nitrogen, which is lossy in the millimeter-wave range. We kept the cold sample free of condensation, a significant source of loss, by enclosing the whole system in a plastic box filled with dry nitrogen gas. The copper is attached to the enclosure by a G10 rod, which provides support and thermal isolation from the enclosure. We measured transmittance of the cooled sample using the same method as for the warm sample except with the alternative sample holder. 

As shown in Figure \ref{fig:3ar}, reflection was suppressed to below 10\% over 92\% and 104\% fractional bandwidth for the two-layer and three-layer coatings, respectively. Cooling the samples reduced the band-integrated absorption loss from 15\% to less than 1\% for the two-layer coating and from 21\% to 10\% for the three-layer coating. The larger loss for the three-layer coating can be attributed to the thicker epoxy layers and high absorption in strontium titanate. However, in typical CMB experiments, lenses operate around 4 Kelvin. Because the loss tangent decreases as function of temperature \cite{Lamb}, we expect a further reduction in loss at operating temperatures. To calculate the bandwidth of low reflection for the three-layer coating, we corrected for this loss and measured the fractional bandwidth above 90\% transmission. This bandwidth is only 12\% greater than the two-layer coating bandwidth. Therefore, for some applications it may be preferable to use the two-layer coating due to its simplicity and lower absorption loss. 

\vspace{-20pt}
\section{Lens Coating}\label{sec:lenscoating}
\begin{figure}[htbp]
\centering
\includegraphics[width=7.4cm]{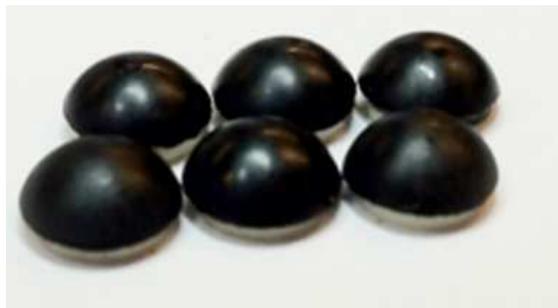}
\vspace{-3pt}
\caption{Photograph of two-layer AR coated alumina lenses with a diameter of 6.35 mm.}
\label{fig:coatedlens}
\end{figure}

To make a sufficiently precise AR coating on a lens, we designed a mold with a cavity that leaves a thin gap between the lens and mold. The cavity was made using a precision machined, ball-ended mill from Pacific Reamer. We were able to create coatings with a tolerance of 25~$\mu\mathrm{m}$, which is approximately $10\%$ of the thickness of each layer. We sprayed the cavity with mold release, and then filled the cavity with the appropriate amount of mixed epoxy that gives high concentricity and an accurate depth. Afterwards, we inserted the lens in the mold and cured the epoxy in a $90^\circ \mathrm{C}$ oven for a few hours. For additional layers, we repeated the process using molds with different spacing. The coated lenses are shown in Figure~\ref{fig:coatedlens}. 

We assessed the quality of the coatings by taking photographs of the side profile and fitting the surface to the expected circular shape. From the fit, we verified the diameter of the coatings was within 25~$\mu\mathrm{m}$ and translation errors were within 25~$\mu\mathrm{m}$ in all directions. We also confirmed the accuracy by removing the AR coating from mold-release coated lenslets and measuring directly with micrometers. Our tolerance corresponds to approximately 10\% of a single layer's thickness, which would result in less than a 1\% decrease in transmittance.

To test cryogenic adhesion, we made twelve two-layer coated 6.35 mm diameter lenses. We kept one sample as a control and slowly cooled nine samples in a vacuum to liquid nitrogen temperature. These nine samples all survived ten slow thermal cycles in the dewar and 18 dunks in liquid nitrogen, returning to room temperature between dunks. Two additional samples were rapidly thermal cycled between room temperature and liquid nitrogen temperature until failure. Failure for one sample occurred after 18 dunks and the other after 50 dunks. There was no change to the control sample that was kept at room temperature. We concluded that the adhesion of the coating is strong enough for our applications. 

\vspace{-20pt}
\section{Conclusion}\label{sec:conclusion}
\vspace{-6mm}
By devising methods to tune the dielectric constant of a mixture between 2.06 and 7.44, we have created effective epoxy-based, broadband anti-reflection coatings for millimeter-wave optics. We reduced the reflection from an alumina slab to less than 10\% over 92\% and 104\% fractional bandwidths with the respective two-layer and three-layer anti-reflection coatings. When samples were cooled to 140 Kelvin, absorptive loss was suppressed to less than 1\% in the two-layer coating and to 10\% in the three-layer coating. Using a precise molding technique, we achieved high precision coating application to a curved surface. We also demonstrated that the coatings can survive numerous thermal cycles.
\vspace{-10pt}
\section*{Acknowledgements}
We thank NASA for its support through grant NNX10AC67G. Praween Siritanasak is supported by the Royal Thai Government fellowship.
\vspace{-10pt}
\end{spacing}

\end{document}